# Growth and development of pure Li$_2$MoO$_4$ crystals for rare event experiment at CUP

**J. K. Son,**[a] **J. S. Choe,**[a] **O. Gileva,**[a] **I. S. Hahn,**[b,c] **W. G. Kang,**[a] **D. Y. Kim,**[a] **G. W. Kim,**[a] **H. J. Kim,**[d] **Y. D. Kim,**[a,e] **C. H. Lee,**[a] **E. K. Lee,**[a] **M. H. Lee,**[a,e,*] **D. S. Leonard,**[a] **H. K. Park,**[f] **S. Y. Park,**[g,a] **S. J. Ra**[a] **and K. A. Shin**[a]

[a] *Center for Underground Physics, Institute for Basic Science (IBS),*
  *Daejeon, 34126, Korea*

[b] *Center for Exotic Nuclear Studies, Institute for Basic Science (IBS),*
  *Daejeon, 34126, Korea*

[c] *Department of Science Education, Ewha womans University,*
  *Seoul, 03760, Korea*

[d] *Department of Physics, Kyungpook National University,*
  *Daegu, 41566, Korea*

[e] *IBS School, University of Science and Technology (UST),*
  *Daejeon, 34113, Korea*

[f] *Department of Accelerator Science, Korea University,*
  *Sejong, 30019, Korea*

[g] *Department of Physics, Ewha Womans University,*
  *Seoul, 03760, Korea*
  *E-mail*: mhlee@ibs.re.kr

ABSTRACT: The Center for Underground Physics (CUP) of the Institute for Basic Science (IBS) is searching for the neutrinoless double-beta decay (0νββ) of $^{100}$Mo in the molybdate crystals of the AMoRE experiment. The experiment requires pure scintillation crystals to minimize internal backgrounds that can affect the 0νββ signal. For the last few years, we have been growing and studying Li$_2$MoO$_4$ crystals in a clean-environment facility to minimize external contamination during the crystal growth. Before growing Li$_2^{100}$MoO$_4$ crystal, we have studied Li$_2^{nat}$MoO$_4$ crystal growth by a conventional Czochralski (CZ) grower. We grew a few different kinds of Li$_2^{nat}$MO$_4$ crystals using different raw materials in a campaign to minimize impurities. We prepared the fused Al$_2$O$_3$ refractories for the growth of ingots. Purities of the grown crystals were measured with high purity germanium detectors and by inductively coupled plasma mass spectrometry. The results show that the Li$_2$MoO$_4$ crystal has purity levels suitable for rare-event experiments. In this study, we present the growth of Li$_2$MoO$_4$ crystals at CUP and their purities.

---

[*] Corresponding author.

# Contents

## 1. Introduction

Neutrinos are among the most abundant particles in the universe, but their natures are not fully understood yet in modern physics. Recent observations of neutrino oscillations by several experiments provide conclusive evidence that neutrinos have non-zero masses. However, absolute masses and other properties of neutrinos are still unknown. These unknowns motivate neutrinoless double-beta (0νββ) decay searches. The Advanced Molybdenum based Rare process Experiment (AMoRE) collaboration searches for the 0νββ decay using scintillation crystals containing the $^{100}$Mo isotope. This isotope is known as one of the most promising isotopes for the 0νββ decay search because of its high transition energy (Q = 3.03 MeV) and relatively high natural abundance (9.6%). The AMoRE experiment aims to resolve unanswered questions related to neutrinos [1-5].

    In recent years, many inorganic scintillation crystals involving molybdenum were studied and developed for 0νββ decay searches. These included CaMoO$_4$, Na$_2$Mo$_2$O$_7$, PbMoO$_4$, ZnMoO$_4$, Cs$_2$Mo$_2$O$_7$, Cs$_3$Mo$_3$O$_{10}$, Li$_2$MoO$_4$ (LMO) crystals, and others. However, each molybdenum crystal has its drawbacks in the context of the AMoRE experiment. First, the CaMoO$_4$ crystal is well known to have the brightest scintillation of these crystals, but the calcium also contains the $^{48}$Ca isotope, which is also a 0νββ decay candidate with a corresponding 2νββ decay that has been observed [6]. $^{48}$Ca has a rather low natural abundance (0.187%), but has a Q-value of 4274 keV, higher than the 3035 keV Q-value of $^{100}$Mo. This high Q-value implies that $^{48}$Ca 2νββ decay can produce background signals at the $^{100}$Mo 0νββ region of interest [7]. Thus, the $^{48}$Ca isotope in calcium must be depleted enough to sufficiently reduce this background. This introduces

significant difficulties for the experiment, not only for an economical reason but also for preparing the raw materials for crystal growth. Cracking issues during crystal growth have been reported for ZnMoO$_4$ and Na$_2$Mo$_2$O$_7$ crystals [8, 9]. For PbMoO$_4$, the low concentration of molybdenum, only 27 wt%, is a disadvantage relative to other crystals. Low molybdenum concentrations imply that a higher total mass is required for a 0νββ search. This also implies that the same total background level requires lower specific activities in the crystals. In the case of Cs$_2$Mo$_2$O$_7$ and Cs$_2$Mo$_3$O$_{10}$, there is a drawback in particle physics experiments because they have the cracking issues and because the light emissions are at relatively high wavelengths within the visible light range. Research on molybdenum crystal scintillators has continued in many directions [10-12].

For the AMoRE experiment, the molybdenum-based crystal needs to perform well on the metrics mentioned above. Even though the Li$_2$MoO$_4$ crystal is reported to have a low intensity of scintillation light compared with others, it can be a good candidate because it has a high concentration of molybdenum (55 wt%) and because this crystal is relatively easy to grow.

## 2. Li$_2$MoO$_4$ crystal growth

### 2.1 Facilities and growth preparation

There are a few techniques for crystal growth such as the Czochralski (CZ), Kyropoulos, and Bridgman methods. The CZ technique is easier than the others. This method has an advantage for controlling the diameter and shape of the crystal. The grown ingot should be a suitable size for machining it into the final detector crystal using methods such as cutting and polishing. This relates directly to material waste, which impacts the experiment budget. Thus, we have studied growing the LMO crystals using the CZ method even though CUP has equipment for all three growth methods mentioned.

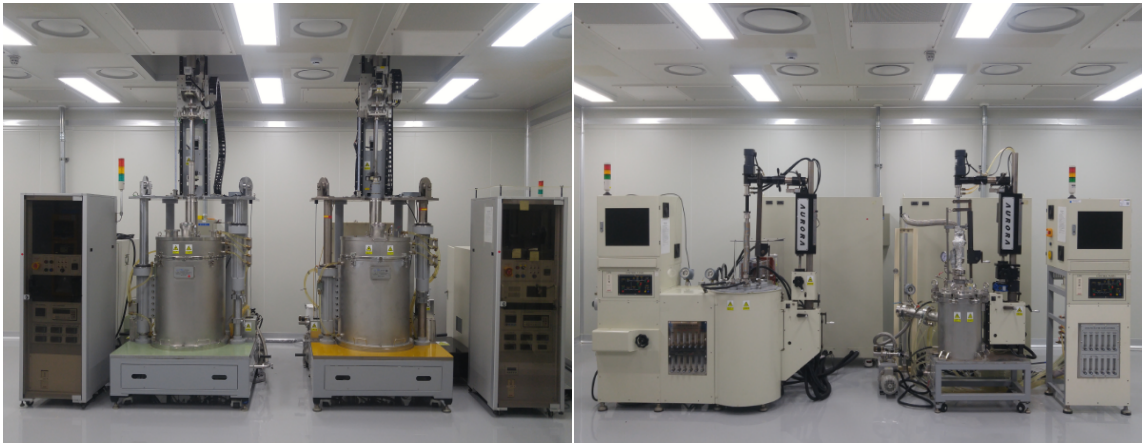

**Figure 1.** Czochralski (left) and Kyropoulos (right) machines in a clean room at the CUP [13].

As shown in Fig 1, we installed the conventional type CZ machines in a class 10,000 cleanroom to produce ultra-pure Mo-based crystals. This room is controlled for low humidity because of the hygroscopic nature of LMO [14]. The air is cleaned with HEPA filters.

We practiced and studied the growth of Li$_2^{nat}$MoO$_4$ as an R&D before growing the Li$_2^{100}$MoO$_4$ crystals to be used in the AMoRE experiment. Li$_2$CO$_3$ and MoO$_3$ powders were mixed with a 1:1 molar ratio as a raw material for the crystal growth. The powder mixture is loaded into the platinum crucible and then is put in the induction work coil at the growth chamber. All devices used in this process are rinsed with 1 ~ 2 vol% of nitric acid solution. The crucible is surrounded by refractories, and the chamber is filled with high purity air (99.999%). The refractories are the internal insulation required for stable growth and were selected from several products to reduce the impurities of the ingot during the growing process. Refractory selection is presented in Sec. 3.3 based on high purity germanium (HPGe) measurements of candidate materials. The mixed powder produces carbon dioxide ($CO_2$) during the synthesis process as the temperature increases. Thus, we slowly flow high purity air to flush $CO_2$ from the chamber.

## 2.2 Growth of ingot

The CZ grower at CUP has a high-resolution load cell to accurately measure the weight of the crystal as the crystal is grown, or pulled, upwards. The ratio of weight change to height change is associated with the diameter and shape of the crystal being grown. Depending on the weight change, the program of the grower can automatically control the heater power. Therefore, the pulling rate and seed rotation speed can be maintained at 2.0 mm/h and 10 rpm respectively. Dimensions of a grown ingot are ~ 50 mm (Ø) × ~140 mm (H) as shown in Fig. 2. We grew several LMO crystals to optimize the growth condition. After growing the crystals, we annealed them for 100 hours in an air atmosphere to release the thermal-stress.

We grew three crystals, which we'll call normal LMO crystals, using commercial raw powder. We also grew five purified LMO crystals from purified powder, and one double-crystalized LMO crystal from the melt of two purified LMO crystals [15, 16]. These crystals were all grown from natural $MoO_3$ powder and denoted as $Li_2^{nat}MoO_4$ crystals. Recently, we have successfully grown $Li_2^{100}MoO_4$ ingots (enriched LMO) using the enriched $^{100}MoO_3$ powder ($^{100}Mo$ isotope ratio > 95%, 99.997% purity grade). These enriched LMO crystals, like the one shown in Fig. 2, can be used as actual detector elements for the AMoRE experiment.

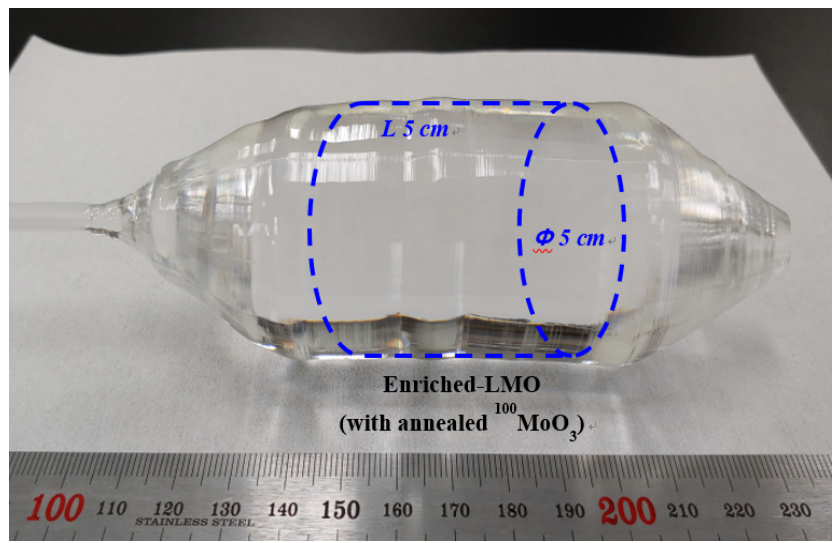

**Figure 2**. A photograph of grown $Li_2^{100}MoO_4$ crystal at the CUP. The dotted lines in the picture show the actual size of the detector element for use in the experiment.

## 3. Radioactivity Assay

### 3.1 Sample preparation

For rare event experiments, the background level is the most important issue. Radioactive impurities in the grown crystals produce what we call internal backgrounds, backgrounds which

– 3 –

cannot be reduced with shielding. Therefore, the growth of ultra-pure $Li_2MoO_4$ crystals is required. We confirmed levels of radioactive contaminants in the crystals by gamma counting with HPGe detectors, and by direct isotopic detection using inductively coupled plasma - mass spectroscopy (ICP-MS). The results are listed specifically in the following sections.

To perform these assays, samples must be prepared from the grown ingots. For HPGe measurements, the top(shoulder) and bottom(tail) of the ingots were removed with saw cuts leaving only the main body. For ICP-MS, small samples were taken from the material directly adjacent to this body section. For several reasons, particularly including the hygroscopic nature of the LMO crystal, special procedures and humidity-controlled equipment were required for sample preparation. Fig. 3 shows a crystal processing machine and a glove box for handling the ingots or samples at the CUP.

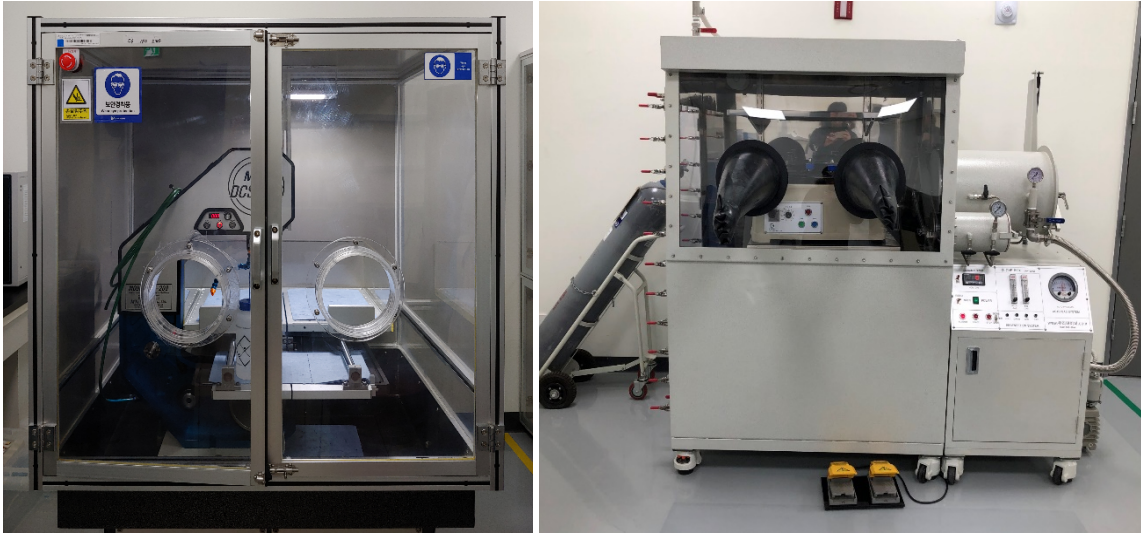

**Figure 3.** A cutting machine that can cut ingot or samples and is equipped with a recovery tool for the cutting residue (left). The glove box system (right) can control humidity level under 0.1 ppm to handle the crystals.

### 3.2 ICP-MS results

### 3.2.1 Raw materials

We have studied $L^{nat}MO$ crystal growth using two different purity grades of $MoO_3$ raw material. One is the commercial $MoO_3$ powder (99.95%, Alfa Aesar) and the other is the highly purified $MoO_3$ powder produced with the sublimation technique at the CUP [17, 18]. For the $Li_2CO_3$ powder, we used the readily available 99.998% purity grade from Alfa Aesar. ICP-MS (Agilent 7900, Agilent Technologies) was used to measure concentrations of contaminant elements such as K, Sr, Ba, Pb, Th, and U. The elements K, Th, and U are radioactive background-producing impurities, and Ba and Pb provide information on general purity and serve as tracers for purification of $^{228}Ra$, $^{226}Ra$ and $^{210}Pb$. Systematic uncertainties for results in Table 1 and Table 2 below are confirmed to be about 10% for Sr, Ba, and Pb, and about 15% for Th and U as previously demonstrated in Ref. [17]. The results in the present work were measured using the same techniques as in Ref. [17] and using the same lots for raw materials. The measurements were however separate measurements taken at later dates with separate samples and different purification batches. The tables thus show somewhat different values compared with the referenced data, but the purity grade is similar. Because of possible variability in the purification process, we used ICP-MS for each crystal to check the purities of the materials used for the growths, in addition to the purities of the grown crystals.

**Table 1.** ICP-MS results of raw materials for L$^{nat}$MO crystal growth (Unit: ppb). *Analysis of K value is the conversion value from HPGe measurement [17].

| Sample | K | Sr | Ba | Pb | Th | U |
|---|---|---|---|---|---|---|
| Li$_2$CO$_3$ powder (99.998%) | 530 | 6.8 | 43.4 | 2.5 | < 0.006 | 0.056 |
| MoO$_3$ powder (99.95%) | *23,403 ±4,774 | 23.7 | 2,829 | 107 | 0.091 | 7 |
| Purified MoO$_3$ powder (@ IBS) | 6,783 | 1.0 | 84.8 | 5.4 | < 0.014 | 0.136 |
| $^{100}$MoO$_3$ powder (Raw powder after annealed) | 348 | 2.2 | 14.9 | 6.2 | 0.064 | 0.083 |

The ICP-MS result for the raw material shows that the concentrations of all elements except Th are approximately ppb level or higher, as summarized in Table 1.

### 3.2.2 Grown crystals

Table 2 summarizes the results of ICP-MS measurements for the L$^{nat}$MO and L$^{100}$MO ingots grown at CUP. In general, each impurity levels of all the LMO crystals were similar.

When comparing the ICP-MS results of the raw materials in Table 1 with the grown LMO crystals in Table 2, the amount of each impurity in the grown LMO has decreased significantly compared to the raw materials. The measured values for elements except K and Ba are lower than the detection limits, which are at the ppt level. This can be interpreted as the effect of segregation during the crystal growth process [19]. As such, we also grew a double-crystalized ingot because crystallization could be considered as one of the most effective methods of purification [16].

**Table 2.** ICP-MS results of grown LMO crystals at IBS (Unit: ppt). Sample types are described in detail in Sec. 2.2.

| Sample Type | K (ppb) | Sr | Ba | Pb | Th | U |
|---|---|---|---|---|---|---|
| Normal LMO | 347 | < 15 | 5,445 | < 300 | < 15 | < 16 |
| Purified LMO | 39 | < 50 | 6,300 | < 100 | < 8 | < 8 |
| Double-crystallized LMO | < 30 | < 50 | 4,744 | < 100 | < 8 | < 8 |
| Enriched LMO | < 60 | < 80 | 1,757 | < 100 | < 8 | < 8 |

However, ICP-MS results for double-crystalized LMO and purified LMO were comparable, with the caveat again that most concentrations were below detection limits for both crystal types. Of the two processes, double-crystallization is significantly more wasteful than powder purification.

### 3.3 HPGe results

#### 3.3.1 Contamination from grower components

Two main components of the CZ chamber can affect crystal contamination during growth. The first is the crucible in which the raw material is loaded, and the second is the refractories surrounding the crucible.

The elements of platinum group metals such as rhodium, iridium, platinum (Pt), and their alloys are usually considered to be safe as common knowledge within the industry, without serious risk of contamination for crystal growth. Because of their chemical stability and oxidation resistance, they are extensively used for high-temperature applications. Although Pt crucibles are not widely used for the growth of oxide materials, due to the relatively low melting point (1768 ˚C), Pt crucibles can be used in an air atmosphere. We attempted to grow $CaMoO_4$ crystals in a Pt crucible. This failed because the Pt crucible was transformed by the temperatures needed for the high melting point of the crystal. Platinum-alloys were also examined, but some elements contained in the alloys are candidates for $0\nu\beta\beta$ decay and can contaminate the crystals during the growth [20-22]. However, we have been using Pt crucibles for LMO crystal growth because the melting point of the LMO crystal (701˚C) is much lower than that of the Pt crucible [16].

HPGe detectors were used to measure the activities of radioactive contaminants in the refractories and the grown crystals. These samples were measured using two coaxial Canberra-brand 100% HPGe detectors, named CC1 and CC2, located at the A5 tunnel in the Yangyang underground laboratory (Y2L) in Korea. The measurement periods for samples were typically for about 5 days. The weights of refractory samples were about 300 ~ 400 g and those of the raw materials were about 2 kg [23-25].

**Table 3**. HPGe results of original and pure refractories. (unit: mBq/kg).

| Sample | Measurement period (Day) | $^{226}Ra$ ($^{238}U$ chain) | $^{40}K$ | $^{228}Ac$ ($^{232}Th$ chain) | $^{228}Th$ ($^{232}Th$ chain) |
|---|---|---|---|---|---|
| Refractory provided by TPS Co. | 4.7 | 2200 ± 200 | 12000 ± 2400 | 4100 ± 590 | 300 ± 270 |
| Fine Tech. Co. Raw material | 4.6 | 21 ± 2.1 | 140 ± 29 | 11 ± 1.9 | 10 ± 1.2 |
| Fine Tech Co. Raw material with additional heat treatment | 4.9 | 38.7 ± 3.7 | 310 ± 63.6 | 21.5 ± 3.6 | 4.3 ± 0.7 |
| Final fused product, from Fine Tech Co. material. | 5.3 | 10.6 ± 1.4 | 640 ± 166 | 10.0 ± 3.0 | 7.3 ± 1.2 |

We measured activities of contaminants in the refractory products provided by TPS Co., the manufacturer of the existing CZ grower, but the result did not satisfy our expectations. Therefore, we measured activities in high-purity raw materials (from Fine Tech Co.) based on $Al_2O_3$. $Al_2O_3$-based refractories are inexpensive shock-resistant materials that could be the chemical resistant to interaction with the melt at high temperatures, making them a promising solution [26-28]. HPGe measurements results for samples of the Fine Tech Co. refractory material with and without extra heat treatment are shown in Table 3 and are compared with the original refractory supplied by TPS Co. The raw material with additional heat treatment was prepared by ourselves to check any changes in raw materials at high temperatures in a normal atmosphere for a long period. From the results, we confirmed that there are not enough increases in impurities between the two samples.

The final fused products supplied by Fine Tech Co. were produced by filling a mold with a mixture of the $Al_2O_3$ raw material (without additional heat treatment) and a binder agent that hardens the mixture. The original refractories by TPS Co. have been also produced similarly, and the extra information of composition or ratio of the mixture could not be obtained from both companies. Activities in this new refractory material were also measured by HPGe, and are also shown in Table 3. The activities measured were orders of magnitude lower than the original materials except for K, and in our judgment were likely low enough to minimize impact to the overall crystal contamination. For reference, assuming chain equilibrium, the activity results reported here correspond to less than 2 ppb concentrations of natural K, Th, and U. Thus, we replaced the refractories for the Czochralski grower at CUP with the fused Fine Tech Co. refractory.

### 3.3.2 Grown crystals

The purity of the grown crystals can be checked not only by ICP-MS but also by gamma counting with HPGe. HPGe is often more practical and/or more sensitive for the measurement of potassium contamination. It also measures different parts of the radioactive decay chains of U and Th, specifically $^{226}Ra$ in the $^{238}U$ chain, and $^{228}Th$ and $^{228}Ac$ in the $^{232}Th$ chain. These lower-chain activities often create larger background concerns than, and are not guaranteed to be in equilibrium with, the upper chain elements measured by ICP-MS. Results of HPGe measurements of grown crystals are shown in Table 4. In the case of ICP-MS, only a small piece of the grown ingot was sampled and measured. However, for HPGe, the entire main body of the ingot was measured, excluding the shoulder and tail parts. The weights of grown ingots were about 600 g, but the body parts, usable in the AMoRE experiments and measured with HPGe, were about 300 ~ 400 g. The measurement periods of each crystal were slightly different and are listed in Table 4. We note that the ingots measured with HPGe were not always the same ingots used for ICP-MS, so the results are representative of the process. The contamination levels of $^{40}K$ found in the natural-LMO crystals correspond to about 1 ppm level of natural potassium.

Similar to ICP-MS, HPGe results of the grown LMOs are also lower than the measurement detection limits for activities from the $^{238}U$ and $^{232}Th$ chains. The corresponding U and Th concentrations, assuming chain equilibrium, are below the level of 1 ppt.

**Table 2.** HPGe results of grown LMO crystals at IBS (unit: mBq/kg). Sample types are described in detail in Sec. 2.2.

| Sample Type | Measurement period (Day) | $^{226}$Ra | $^{40}$K | $^{228}$Ac | $^{228}$Th |
|---|---|---|---|---|---|
| Normal LMO | 17.3 | < 3.3 | 28.9 ± 9.7 | < 2.6 | < 1.5 |
| Purified LMO | 21.5 | < 0.6 | < 8.2 | < 2.1 | < 1.0 |
| Double-crystallized LMO | 12.8 | < 1.2 | < 13.8 | < 3.2 | < 1.3 |
| Enriched-LMO | 27.7 | < 1.6 | <13.7 | < 5.7 | < 3.4 |

## 4. Conclusion

$Li_2MoO_4$ is a candidate scintillation crystal for the detection of neutrinoless double-beta decay. At CUP in IBS, we have successfully performed test growths of $Li_2^{nat}MoO_4$ and $Li_2^{100}MoO_4$ crystals using the conventional Czochralski method. The dimensions of the grown crystals are about 50 mm (Ø) × 140 mm (H). $^{Nat}$Mo-based crystals were grown using commercial $MoO_3$, highly purified $MoO_3$ produced at CUP by sublimation, and by re-melting two previously grown LMO crystals with purified $MoO_3$ as the raw material. We have also tried to minimize impurities introduced during the growth process. We selected refractories of fused $Al_2O_3$ materials for use in the crystal grower based on HPGe impurity measurements.

Based on the results of ICP-MS and HPGe, we can see that the grown LMO crystals are very pure. Also, after replacing the refractories, there is a significant reduction of contamination from the crystal growth process itself. Thus, our grown LMO crystals can be a suitable candidate crystal for rare event search experiments such as AMoRE.

### Acknowledgments

This work was supported by the Institute for Basic Science and the National Research Foundation of Korea funded by the Ministry of Science and Technology, Korea (Under Grant: IBS-R016-D1, 2018R1A2A1A05022079, and 2018R1A6A1A06024970).